\documentstyle[12pt,prl,aps]{revtex}

\draft
\begin{document}
\title{
Exact Diagonalization Approach for the $D=\infty$ Hubbard Model}
\author{Michel Caffarel$^{*,+}$ and Werner Krauth$^{**}$}

\address{
$^{*}$ CNRS-Laboratoire de Physique Quantique$^1$\\
IRSAMC, Universit\'{e} Paul Sabatier\\
118, route de Narbonne; F-31062 Toulouse Cedex; France\\
e-mail: mc@tolosa.ups-tlse.fr\\
$^{**}$ CNRS-Laboratoire de Physique Statistique de l'ENS\\
24, rue Lhomond; 75231 Paris Cedex 05; France\\
e-mail: krauth@physique.ens.fr\\}
\date{June 1993}
\maketitle
\begin{abstract}
We present a powerful method for calculating the thermodynamic properties
of the Hubbard model in infinite dimensions, using an exact
diagonalization of an Anderson model with a finite number of sites.
At finite temperatures, the explicit diagonalization of the Anderson
Hamiltonian allows the calculation of Green's functions
with a resolution far
superior to that of  Quantum Monte Carlo calculations.
At zero temperature, the Lancz\`os method is used
and yields the essentially exact zero-temperature solution of the model,
except in
a region of very small frequencies.
Numerical results
for the half-filled case in the paramagnetic phase (quasi-particle
weight, self-energy, and also real-frequency spectral densities)
are presented.
\end{abstract}
\newpage
Following the pioneering work of Metzner and Vollhardt \cite{MV}, the limit of
large dimensions for models of
strongly correlated fermions has received much
attention. In this limit, the highly intricate quantum many-body
problem simplifies considerably and leads to a non-trivial mean-field
theory \cite{VOL}. Remarkably, this limit
captures many features of the physics in finite dimensions
and gives a very successful description of quantum fluctuations.

In spite of the considerable simplification obtained in taking the
large D limit, the mean-field equations still have to be
solved numerically. Up to now, all calculations \cite{JAR}, \cite{RZK},
\cite{GKr1} have relied on
the Hirsch-Fye Quantum Monte Carlo (QMC)
algorithm \cite{HF}. A major limitation of this scheme
is the difficulty of accessing the low-temperature regime, where
statistical and
finite time-step discretization  errors of the QMC algorithm
become very important.

In this paper, we present a powerful method for solving these mean-field
equations, which leads to an essentially exact solution in the
imaginary frequency domain. As an example, we consider the Hubbard model
on a lattice of infinite connectivity $z \rightarrow \infty$ which,
after proper rescaling of the kinetic energy, is written as
\begin{equation}
H = - \sum_{<ij> \sigma} \frac{1}{\sqrt{2z}} c_i^+ c_j + h.c. +
U \sum_i n_{i\uparrow} n_{i\downarrow},
\label{Hhubb}
\end{equation}
The calculation of the single-site properties of the Hubbard model
in this limit reduces to the self-consistent determination of
the on-site Green's
function $G(\omega)$ of the Hubbard model and of a bath Green's function
$G_0(\omega)$, which describes the interaction on the single site
with the external environment.
$G(\omega)$ and $G_0(\omega)$
are related by a self-consistency condition which, on the Bethe lattice, reads:
\begin{equation}
G_0^{-1}(\omega) = \omega +\mu -\frac{1}{2} G(\omega)
\label{sc}
\end{equation}
It is for simplicity only that we restrict our attention in this paper
to the $z\rightarrow \infty$ Bethe lattice.

As is well known \cite{VDV} \cite{GKo},
the  on-site Green's function of the Hubbard model may be interpreted as
the  Green's function of
an Anderson model
\begin{equation}
H_{And} = \epsilon_d\sum_{\sigma}d^+_{\sigma}
d_{\sigma} +
\sum_{\sigma,k=2}^{n_s} \epsilon_k  a^+_{k \sigma}
a_{k \sigma}
 + U n_{d\uparrow} n_{d\downarrow}
+ \sum_{\sigma, k=2}^{n_s} (V_{k} a^+_{k \sigma}
d_{\sigma} + h. c. )
\label{And}
\end{equation}
in which the function $G_0(i\omega_n)$ is given by the $U=0$ Green's
function of the impurity
\begin{equation}
G_0(i\omega_n)=G_0^{And}(i\omega_n) = [i\omega_n -\epsilon_d - \mu-
\sum_{k=2}^{n_s}\frac{ V_{k}^2 }
{i\omega_n -  \epsilon_k}]^{-1}
\label{G_0}
\end{equation}
Given the infinite number of degrees of freedom of the models
defined in eq. (\ref{Hhubb}) and eq. (\ref{And}), it is evident that
strict self-consistency can only be obtained with a
continuous Anderson model, {\it i. e.} with $n_s = \infty$. The main
result of the present paper is that a systematic
approximation of $G_0(i\omega)$ with a finite-$n_s$ Anderson model
gives extremely good results. We stress from the beginning that we
are interested in an approximation of the imaginary-frequency Green's
functions only.

In practice, we approximate any
$G_0^{-1}(i\omega)$ by a function $G_0^{-1\;And}(i\omega)$
with a finite number $n_s$ of sites. This can be cast into a
minimization problem in the variables
$\epsilon_k$ and
$V_{k}$.
For this paper, we choose the following cost function:
\begin{equation}
\chi^2 = \frac{1}{n_{max} +1} \sum_{n=0}^{n_{max}} {|G_0^{-1}(i\omega_n)
- G_0^{-1\;And}(i\omega_n)|}^2
\label{chisq}
\end{equation}
where $n_{max}$ is chosen sufficiently large
($\omega_{nmax} >> max_{k}(\epsilon_{k})$) \cite{footchi}.
We search for the parameters $\epsilon_k$ and
$V_{k}$ minimizing the $\chi^2$ in eq. (\ref{chisq}) with a standard
conjugate gradient method \cite{footsymm}.

For a small number of sites, $n_s\leq 6$,
the Green's function $G(i\omega_n)$ can
be obtained exactly from the complete set of
eigenvectors and eigenvalues of the Anderson Hamiltonian
eq. (\ref{And}).
The procedure
\begin{equation}
G_0^{-1}(i\omega_n)\stackrel{eq.\;(\ref{chisq})}{\longrightarrow}
G_0^{-1\; And}(i\omega_n)
\stackrel{eq.\;(\ref{And})}{\longrightarrow}G(i\omega_n)
\stackrel{eq.\;(\ref{sc})}{\longrightarrow}G_0^{-1}(i\omega_n)
\label{loop}
\end{equation}
is then iterated to convergence.

The following observations are made:

1) We notice in general very small differences between
$G_0^{-1}(i\omega)$ and $G_0^{-1\;And}(i\omega)$
as expressed by small minimal values of $\chi^2$ in eq. (\ref{chisq}).
$\chi^2$ decreases by approximately a constant factor each time we add
one more site.

2) The extensive comparisons with QMC results \cite{GKr1} which we have
undertaken
indicate that exact diagonalization is by far the superior method for this
problem.
As an example, we show in fig. \ref{gtau} QMC and exact diagonalization data
for the half-filled Hubbard model at $\beta=32$ and $U=3$. The Monte
Carlo data are shown for a imaginary-time
discretization of $\Delta{\tau} = 1, .5$, and .25 ({\it cf, e. g.}
\cite{GKr1}),
and the exact diagonalization data for $n_s = 3,5$.
It may be worthy of notice that the diagonalization
calculations can be obtained in a few minutes on a work station, while
for the  QMC data acquisition (at $\Delta{\tau} = 0.5$)
several hours were needed (several days for $\Delta{\tau} = 0.25$).

Beyond $n_s = 6$, the size of the Hilbert space becomes too large
for an explicit diagonalization of the  Anderson Hamiltonian. However,
the calculation of {\it zero-temperature} Green's functions is still
possible by means of the
Lancz\`os algorithm \cite{LAN}, which allows us to easily calculate
$G(i \omega)$ and $G_0(i \omega)$ up to  $n_s \sim 10$ on a
workstation \cite{footcomp}.
The fit with the Anderson
model is performed as before. We simply replace
the Matsubara frequencies
by a fine grid of imaginary frequencies, which correspond to a
``fictitious" inverse temperature $\beta$
($\omega_n= (2n+1)\pi/\beta$).
$\beta$ introduces a low-frequency cutoff in an obvious way.
In fig. \ref{gomega}
we display the functions $G_0^{-1}(i\omega)$ and $G_0^{-1\;And}(i\omega)$
for $U=2$ and $U=4.8$.
At the scale of the figure the two curves can hardly be distinguished,
and an essentially
perfect fit ({\it i.e.} perfect self-consistency)  is obtained in the
whole range of frequencies. The inset in the figure shows a blow-up of
the small frequency regime as the number of sites is increased. Notice
the {\it systematic} amelioration of the fit.
Furthermore, the 'physical' Green's function
$G_0^{-1}$ is extremely independent of $n_s$, especially at high
frequency.
Already at $\omega=0.11$, {\it e.g.}, $G_0^{-1}$ varies by less than
$0.0001$ between $n_s=6,8,$ and $10$.

For the  data at $U=4.8$, the quality of the fit is excellent even
with a small number of sites. This is easily
explained by the existence of a physical
cutoff in frequency, which  results from the Mott gap.

We now pass to the calculation of other physical quantities and present in
fig. \ref{quasi} some results for the quasi-particle
spectral
weight Z calculated from the slope of the self-energy $\Sigma =
G_0^{-1} - G^{-1}$.  In the inset of fig. \ref{quasi} we present the
raw data of $Im{\Sigma}(i \omega)$ at small
frequencies from which the
spectral weight is extracted
($Im{\Sigma}(i\omega) \sim  (1-1/Z) \omega +\dots$).
To get a truly stabilized slope of $\Sigma$
we have found it to be  necessary to reach
very low temperatures. The main plot compares
the results at $n_s=10$ with the
``iterated perturbation theory" (IPT) result. This
method is based essentially on the use of a weak coupling calculation
to second-order in $U$ of $\Sigma$ and  has shown to give a satisfactory
interpolation between the small and large $U$ limits (exclusively
at half filling and in the paramagnetic phase) \cite{GKo},
\cite{GKr2},\cite{ZRK}. On a few points
we give in addition the results of the exact diagonalization
at $n_s=6$ and $n_s=8$. Given the extremely good agreement between the
values of $Z$ calculated with $n_s = 8$ and $10$,
we are very confident of the numerical values presented.

As
discussed in ref \cite{GKr2}, the IPT approximation leads to a
first-order Mott-Hubbard transition ({\it cf} fig. \ref{quasi}),
and the quasi-particle
weight $Z$ jumps discontinuously at $U\sim 3.6$.
We have only found limited evidence for such a scenario within the
present approach. At $n_s=6$, we are unable to stabilize two solutions
at the same values of the physical parameters (the coexistence
of two solutions is indicative of a first-order phase transition).
At $n_s=8$, and using a fictitious temperature of $\beta=120$, we
find a coexistence region
within a very small interval of $U$: $4.45 \leq U \leq 4.60$ \cite{trans}.
Even though the question of the order of the transition will have to await
a more detailed investigation, it seems to us to be difficult to
reconcile our numerical results with a abrupt transition anywhere
close to $U=3.6$.

Finally, we show some data concerning
the one-particle spectral densities $\rho (\omega) = -
Im G(\omega + i\epsilon)/\pi$ as obtained from the
Lancz\`os calculation together with IPT-approximation solutions
\cite{GKr2}.
Fig. \ref{dens}
shows the spectral density ($n_s$=10) for different values of $U$. In
the Fermi-liquid regime the spectrum of our finite-size Anderson model
consists of a large number of peaks, while in the
insulating phase we systematically observe a simpler structure made of only
a few peaks. As $U$ is increased we see that $\rho(\omega)$ develops three
well-separated structures: a central quasi-particle feature and two
broad high-energy satellite features corresponding to the formation
of the upper Hubbard band. At $U=4.8$ a gap is observed in good
agreement with the approximate IPT solution. In the insets of Fig. \ref{dens}
we also present the integrated single particle density of states
corresponding to Lancz\`os
and IPT solutions. The agreement between both curves is seen to be very good,
provided we average over a frequency interval of $\omega \sim 0.5$. This
indicates that the calculated spectral density contains
coarse-grained information about the exact solution, as it should be.
Due to the discrete nature of the Anderson model used, the fine details
of the spectrum are poorly reproduced.

A remark is in order here: As is well known, the
continuation of numerical data from the imaginary axis onto the
real-frequency domain is a very difficult problem and constitutes
for example one of the major limitations of the QMC method.
Here we encountered the analytic continuation problem in the 'easy'
direction.
Indeed, in the present work very precise
imaginary frequency data ({\it cf} fig, \ref{gtau}, fig. \ref{gomega})
can be obtained with a representation in
$\omega$ , which very cleary has its limits ({\it cf} fig. \ref{dens}).
It is crucial in this context
that we only attempt to satisfy the self-consistency condition on the
imaginary axis.

In conclusion, we have presented a powerful numerical method for
simulating the $D=\infty$ models for strongly correlated fermions
based on a self-consistent single-impurity model treated by exact
diagonalization. At the temperatures reachable by quantum Monte Carlo
calculations we get essentially the exact solution of the model. At
lower temperatures unreachable by QMC, we get also an amazingly good
solution, except in the region of very small frequencies where some
difficulties appear due to the
finite degrees of freedom in the representation of
the free propagator ($G_0^{-1\;And}$) of the impurity.

Elsewhere \cite{CK} we present
a study of the instability of the
normal phase with respect to superconductivity  of an infinite-D
 two-band Hubbard model
\cite{GKK}. That work and additional calculations on the Hubbard model
away from half-filling clearly show that the exact diagonalization method
presented in this work is in no way limited to the particle-hole symmetric
point of the Hubbard model. Broken-symmetry phases, magnetic fields, {\it etc},
as well as the calculation of susceptibilities \cite{CGK} can be easily handled
within
this approach, which we expect to rapidly become a standard
tool for the investigation of $D=\infty$ systems.

\acknowledgments
We acknowledge helpful discussions with
J. Bellissard, A. Georges, G. Kotliar, D. Poilblanc and T. Ziman.
This work was supported by DRET contract $n^o 921479$.\\
\\
$^{+}$Permanent address:
Laboratoire Dynamique des Interactions Mol\'{e}culaires, Tour 22
Universit\'{e} Paris VI; 4 place Jussieu F-75252 Paris Cedex 05, France

\newpage
{\bf Figure Captions}
\begin{enumerate}
\item{}
\label{gtau}
Comparison of $G(\tau)$ for $U=3$, and $\beta=32$ between exact diagonalization
($n_s=3,5$ (bottom solid lines) the results cannot be distinguished on
the scale plotted)
and Quantum Monte Carlo (solid: $\Delta{\tau}=1$, dashed:
$\Delta{\tau}=0.5$).
Inset: $G(\tau=4)$ {\it vs.}
$\Delta{\tau}^2$ (the QMC algorithm converges roughly in
${\Delta{\tau}}^2$). At $\Delta\tau=0.$ exact diagonalization results for
$n_s=3,5$.
\item{}
\label{gomega}
Plot of $G_0^{-1}(i\omega)$ and of $G_0^{-1\; And}$ for two values of the
interaction, $U=2$ (Fermi liquid-regime) and $U=4.8$ (Mott insulator
regime) at $n_s=10$. The inset gives $G_0(i\omega)^{-1}$ and $G_0^{-1\; And}$
at small frequency for $n_s=6,8,10$. Note the systematic improvement of the
solution. The misfit between the two functions is the only source of error of
the exact diagonalization approach.
\item{}
\label{quasi}
Quasi-particle weight $Z$ as a function of $U$ for the half-filled
Hubbard model. The curve gives the IPT approximation, which predicts
a first-order transition. The crosses give the results for $n_s=10$, with
the corresponding results for $n_s = 6, 8$ at two points.
The inset shows the small-$\omega$ behavior of
$Im{\Sigma}(i\omega)$ for $n_s=6,8,10$
from which the quasi-particle weight
is calculated. Note the excellent convergence with $n_s$.
\item{}
\label{dens}
Density of states $\rho(\omega)$ for different values of $U$ (a value of
$\epsilon=0.01$ is used).
We compare with the IPT density of states \cite{GKr2}. Insets: Comparison
of the integrated densities of states between exact diagonalization and IPT.
\end{enumerate}

\begin{references}
\bibitem{footsymm}
In order to force the particle-hole
symmetry a symmetric distribution of the $\epsilon_l$'s around zero with
one energy kept fixed at zero has been chosen ($n_s-1$ independent
parameters) in most of our simulations.
\bibitem{footcomp}
Higher $n_s$ (say, up to $n_s \sim 16$) could be handled on a
supercomputer using efficiently designed algorithms (that was not
the purpose of this work). In any case,
this would not change quantitatively our well-converged results
(see below).
\bibitem{trans}
Convergence is very slow in the transition region.
There it may take several hundred iterations to destroy an apparently  stable
Fermi-liquid or Mott insulating phase. However, we always
end up with a fully stabilized solution. Such a situation is almost
impossible to handle correctly in expensive QMC calculations at
low temperatures.
\end{references}
\end{document}